\documentstyle[preprint,prc,aps]{revtex}
  \def\vec#1{\mbox{\boldmath $#1$}}
  \def\gsl#1{\rlap{\slash}#1} 
  \def\Disc.{{\rm Im}}
  \def\Cont.{{\rm Re}}
  \def\MN{m}
  \def\Eq.#1{Eq.~(\ref{#1})}

\title{A model-independent study of the QCD sum rule for the $\pi NN$ coupling constant}
\author{Yoshihiko Kondo\footnote{kondo@kokugakuin.ac.jp}}
\address{Kokugakuin University, Higashi, Shibuya, Tokyo 150-8440, Japan}
\author{Osamu Morimatsu\footnote{osamu.morimatsu@kek.jp}}
\address{Institute of Particle and Nuclear Studies, High Energy Accelerator Research Organization, Tukuba, Ibaragi 305-0801, Japan}

\begin{document}
\draft
\maketitle

\begin{abstract}
We reinvestigate the QCD sum rule for the $\pi NN$ coupling constant, $g$, starting from the vacuum-to-pion matrix element of the correlation function of the interpolating fields of two nucleons.
We study in detail the physical content of the correlation function without referring to the effective theory.
We consider the invariant correlation functions by splitting the correlation function into different Dirac structures.
We show that the coefficients of the double-pole terms are proportional to $g$ but that the coefficients of the single-pole terms are not determined by $g$.
In the chiral limit the single-pole terms as well as the continuum terms are ill defined in the dispersion integral.
Therefore, the use of naive QCD sum rules obtained from the invariant correlation functions is not justified.
A possible procedure to avoid this difficulty is discussed.
\end{abstract}
\vspace{12pt}

The QCD sum rule invented by Shifman, Vainshtein and Zakharov
provides us with a way to relate the physical quantities of the hadrons to the matrix elements of the quark gluon composite operators by means of the operator product expansion (OPE)~[\ref{SVZ},\ref{RRY}].

There has already been quite a history in the study of the meson-baryon-baryon coupling constants within the framework of the QCD sum rule~[\ref{RRY}-\ref{KLO}]. 
Two ways of the formulation exist in constructing sum rules:
one is to start from the vacuum-to-vacuum matrix element of the correlation function of the interpolating fields of two baryons and one meson and the other is to start from the vacuum-to-pion matrix element of the correlation function of the interpolating fields of two baryons.
In the pionnering work by Reinders, Rubinstein and Yazaki the former approach was taken but in the following works including the one by the same authors the latter approach was employed.
This is because the former approach has to assume additional extrapolation from the large space-like momentum to the zero momentum for the pion compared to the latter approach.
Furthermore it was shown in Ref.~[\ref{Maltman}] that the three-point function method for the treatment of both the isospin conserving and isospin violating pion-nucleon-nucleon couplings is plagued by problems with higher resonance contamination. 

Therefore, we only discuss here the formulation with the vacuum-to-pion matrix element of the correlation function of interpolating fields of two baryons:
\begin{eqnarray}\label{Pi}
\Pi(p,k)=
-i\int d^4xe^{ipx}\langle 0|T[\eta(x)\bar\eta(0)]|\pi(k)\rangle,
\end{eqnarray}
where $\eta(x)$ is the interpolating field for the nucleon, $|\pi(k)\rangle$ is the pion state with momentum $k$ normalized as $\langle\pi(k')|\pi(k)\rangle=2k_0(2\pi)^3\delta^3(\vec k'-\vec k)$ and the isospin is neglected for simplicity.
The correlation function can be split into four Dirac structures as,
\begin{eqnarray}\label{FDS}
  \Pi(p,k)=i\gamma_5\Pi_1(p^2,pk)+i\gamma_5{\gsl p\over \MN}\Pi_2(p^2,pk)+i\gamma_5{\gsl k\over \MN}\Pi_3(p^2,pk)-i\gamma_5{i\sigma_{\mu\nu}p^\mu k^\nu\over \MN^2}\Pi_4(p^2,pk),
\end{eqnarray}
where $\MN$ denotes the nucleon mass, $\Pi_i$ $(i=1\sim 4)$ are functions of Lorentz invariants, $p^2$ and $pk$, and are called invariant correlation functions.
Reinders, Rubinstein and Yazaki constructed a sum rule for the pion-nucleon-nucleon ($\pi NN$) coupling constant in the chiral limit with $k=0$ from the invariant correlation function with the Dirac structure $i\gamma_5$, $\Pi_1$, in the leading order of the OPE.
They showed that if the sum rule is divided by the odd-dimensional sum rule for the nucleon mass, the result is consistent with the Goldberger-Treiman relation with $g_A=1$, where $g_A$ is the axial charge of the nucleon~[\ref{RRY}].
Shiomi and Hatsuda improved the sum rule by taking into account higher dimensional terms and $\alpha_s$ corrections of the OPE and also the continuum contributions~[\ref{S&H}].
They showed that even after these corrections are taken into account, the Goldberger-Treiman relation with $g_A=1$ holds as long as the same continuum threshold is taken in the sum rule for the $\pi NN$ coupling constant and the odd-dimensional sum rule for the nucleon mass.
In Ref.~[\ref{B&K}] Birse and Krippa pointed out that in the chiral limit the vacuum-to-pion correlation function is obtained just by chirally rotating the vacuum-to-vacuum correlation function and therefore it is obvious that the ratio of the two sum rules is consistent with the Goldberger-Treiman relation, which is just a consequence of the chiral symmetry.
Then, they tried to obtain a new sum rule for the $\pi NN$ coupling constant which is not just a consequence of the chiral symmetry.
They considered the invariant correlation function with the structure $i\gamma_5\gsl k$ and obtained a sum rule by taking the limit, $k=0$, after removing $i\gamma_5\gsl k$.
In all of these works the sum rule is constructed by making an \lq\lq ansatz \rq\rq for the absorptive part of the correlation function based on the effective lagrangian with the pseudoscalar coupling scheme.
In Ref.~[\ref{KLO}] Kim, Lee and Oka examined how the choices of the effective lagrangian, i.e. the pseudoscalar and pseudovector coupling schemes make differences in the sum rules for all the four Dirac structures.
They concluded that only the invariant correlation function with the Dirac structure $\gamma_5\sigma_{\mu\nu}p^\mu k^\nu$, is independent of the two coupling schemes and obtained a sum rule by taking the limit, $k=0$, after removing $\gamma_5\sigma_{\mu\nu}p^\mu k^\nu$.
They also claimed that their sum rule suffers from less uncertainties due to QCD parameters.

Even though we agree with Kim et al. in the point that the sum rule should not depend on the choice of the coupling scheme, we find their argument unsatisfactory because they still rely on particular effective lagrangians.
Actually, we can identify the physical structure of the correlation function just by general principles without referring to effective lagrangian~[\ref{K&M}].

The purpose of the present paper is to reinvestigate the QCD sum rule for the $\pi NN$ coupling constant by applying the procedure of Ref.~[\ref{K&M}].

We clarify the physical contents of the correlation function defined by \Eq.{Pi} without referring to the effective theory.
The correlation function $\Pi(p,k)$ in \Eq.{Pi} has a pole at $p^2=\MN^2$ and $(p-k)^2=\MN^2$ where $p$ and $p-k$, respectively, become on-shell momenta for the nucleon.
The $\pi NN$ coupling constant, $g$, is defined through the coefficient of the pole as
\begin{eqnarray}\label{gpiNN}
\bar u(\vec pr)(\gsl p-\MN)\Pi(p,k)(\gsl p-\gsl k-\MN)u(\vec qs)|_{p^2=\MN^2,(p-k)^2=\MN^2}=-ig\bar u(\vec pr)\gamma_5u(\vec qs),
\end{eqnarray}
where $q=p-k$, $u(\vec pr)$ is a Dirac spinor with momentum $p$, spin $r$ and is normalized as $\bar u(\vec pr)u(\vec pr)=2\MN$ and the unnormalized nucleon interpolating field, $\eta$, is replaced by the normalized nucleon field, $\psi$.

In addition to the pole singularity, $\Pi(p,k)$ has a branch cut singularity starting from the threshold of the pion-nucleon channel, $p^2=(\MN+m_\pi)^2$, to infinity where $m_\pi$ is the pion mass.
We restrict ourselves to the positive energy region, $p_0>0$, for simplicity.

In order to classify the singularity of $\Pi(p,k)$,
we define the vertex function, $\Gamma(p,q',k')$ $(p=q'+k')$, and the pion-nucleon T-matrix, $T(q',k',q,k)$ $(q'+k'=q+k)$, by
\begin{eqnarray}
&&\Gamma(p,q',k')=(\gsl p-\MN)[-i\int d^4xe^{ipx}\langle 0|T[\psi(x)\bar\psi(0)]|\pi(k')\rangle](\gsl q'-\MN),\cr
&&T(q',k',q,k)=(\gsl q'-\MN)[-i\int d^4xe^{iq'x}\langle\pi(k')|T[\psi(x)\bar\psi(0)]|\pi(k)\rangle](\gsl q-\MN).
\end{eqnarray}
The correlation function is related to the vertex function as
\begin{eqnarray*}
\Pi(p,k)={\gsl p+\MN \over p^2-\MN^2}\Gamma(p,p-k,k){\gsl p-\gsl k+\MN \over (p-k)^2-\MN^2}.
\end{eqnarray*}

Let us regard $\Pi(p,k)$ as a function of the center-of-mass energy, i.e. $p_0$ in the frame $\vec p=0$, and consider the discontinuity of the correlation function.
We adopt the following notation for the dispersive (continuous) part and the absorptive (discontinuous) part, respectively:
\begin{eqnarray}
&&\Cont.F(p)\equiv{1\over2}[F(p)|_{p^0=p^0+i\eta}+F(p)|_{p^0=p^0-i\eta}],\cr
&&\Disc.F(p)\equiv{1\over2i}[F(p)|_{p^0=p^0+i\eta}-F(p)|_{p^0=p^0-i\eta}].
\end{eqnarray}
The absorptive part of the correlation function, ${\rm Im}\Pi(p,k)$, can be written as
\begin{eqnarray}\label{TwoParts}
{\rm Im}\Pi(p,k)
=(\gsl p+\MN)\bigg\{&&{\rm Im}{1 \over (p^2-\MN^2)((p-k)^2-\MN^2)}{\rm Re}\Gamma(p,p-k,k)\cr
& &+{\rm Re}{1 \over (p^2-\MN^2)((p-k)^2-\MN^2)}{\rm Im}\Gamma(p,p-k,k)\bigg\}
(\gsl p-\gsl k+\MN).
\end{eqnarray}
The first term represents the pole contribution and the second term represents the continuum contribution.

When the c.m. energy, $p_0$, is above the threshold of the pion-nuclon channel, $\MN+m_{\pi}$, but below that of the next channel, only the pion-nucleon channel contributes in the intermediate states and the absorptive part of the vertex function, ${\rm Im}\Gamma(p,p-k,k)$, is given by
\begin{eqnarray}\label{O_T}
\Disc.\Gamma(p,p-k,k)
&=&-{\pi\over4}{t\over p_0}\int {d\Omega'\over(2\pi)^3}
{\Gamma(p,q',k')}(\gsl q'+\MN){T^*(q',k',q,k)},
\end{eqnarray}
where $t$ is defined through $p_0=\sqrt{\MN^2+t^2}+\sqrt{m_{\pi}^2+t^2}$.
$q'$ and $k'$ are the on-shell momenta for the nucleon and the pion, respectively.

From \Eq.{O_T}, one sees that in the vicinity of the pion-nucleon threshold $\Gamma$ can be expanded as
\begin{eqnarray}\label{ImVF}
\begin{array}{l}
  {\rm Im}\Gamma(p,p-k,k)=\left\{
    \begin{array}{l l}
      0&\quad(p_0< \MN+m_\pi)\cr
      G_1t
      + O(t^3)&\quad(p_0> \MN+m_\pi)
    \end{array}
\right. ,\cr
  {\rm Re}\Gamma(p,p-k,k)=\left\{
    \begin{array}{l l}
      G_0+G_1\tau+O(\tau^3)&\quad(p_0< \MN+m_\pi)\cr
      G_0+O(t^2)&\quad(p_0>  \MN+m_\pi)
    \end{array}
\right. ,
\end{array}
\end{eqnarray}
where $\tau=it$.
In \Eq.{ImVF} the behaviour of the dispersive part is determined from that of the absorptive part by the Cauchy-Riemann equations. 
From \Eq.{ImVF} one sees that
\begin{eqnarray}\label{ReVF}
\left.{\partial\over\partial p_0}{\rm Re}\Gamma(p,p-k,k)\right|_{p_0=\MN+m_\pi-\epsilon}&=&\left\{
    \begin{array}{l l}
      \infty&\quad(m_\pi\not=0)\cr
      \hbox{finite constant}&\quad(m_\pi=0)
    \end{array}\right. ,\cr
\left.{\partial\over\partial p_0}
{\rm Re}\Gamma(p,p-k,k)\right|_{p_0=\MN+m_\pi+\epsilon}&=&0,
\end{eqnarray}
which is crucial in the following discussions.

We discuss the physical content of the invariant correlation function.
Similarly to \Eq.{FDS} we split the vertex function into four Dirac structures as
\begin{eqnarray*}
\Gamma(p,p-k,k)=i\gamma_5\Gamma_1(p^2,pk)+i\gamma_5{\gsl p\over \MN}\Gamma_2(p^2,pk)+i\gamma_5{\gsl k\over \MN}\Gamma_3(p^2,pk)-i\gamma_5{i\sigma_{\mu\nu}p^\mu k^\nu\over \MN^2}\Gamma_4(p^2,pk),
\end{eqnarray*}
where $\Gamma_1 \sim \Gamma_4$ are the invariant vertex functions.

The invariant correlation functions are related to the invariant vertex functions as
\begin{eqnarray}\label{ICFs}
\Pi_1
&=&{{m_\pi^2\over 2}\left(\Gamma_1-\Gamma_2-2\Gamma_3+\left(2-{m_\pi^2\over 2\MN^2}\right)\Gamma_4\right)\over(p^2-\MN^2)((p-k)^2-\MN^2)}
-{{1\over 2}\left(\Gamma_1+\Gamma_2-{m_\pi^2\over \MN^2}\Gamma_4\right)+{1\over 4\MN^2}(p^2-\MN^2)\Gamma_4\over (p-k)^2-\MN^2}\cr
&&-{{1\over 2}\left(\Gamma_1-\Gamma_2-{m_\pi^2\over \MN^2}\Gamma_4\right)+{1\over 4\MN^2}((p-k)^2-\MN^2)\Gamma_4\over p^2-\MN^2}
+{1\over 2\MN^2}\Gamma_4,
\cr 
\Pi_2
&=&{\Gamma_3-\Gamma_4\over p^2-\MN^2}-{\Gamma_2+\Gamma_3-\Gamma_4\over (p-k)^2-\MN^2},
\cr 
\Pi_3
&=&{\MN^2\left(-\Gamma_1+\Gamma_2+2\Gamma_3-\left(2-{m_\pi^2\over 2\MN^2}\right)\Gamma_4\right)\over(p^2-\MN^2)((p-k)^2-\MN^2)}
+{\Gamma_2+\Gamma_3-{3\over 2}\Gamma_4\over (p-k)^2-\MN^2}-{{1\over 2}\Gamma_4\over p^2-\MN^2},
\cr 
\Pi_4
&=&{\MN^2\left(\Gamma_1-\Gamma_2-2\Gamma_3+\left(2-{m_\pi^2\over 2\MN^2}\right)\Gamma_4\right)\over(p^2-\MN^2)((p-k)^2-\MN^2)}
+{{1\over 2}\Gamma_4\over (p-k)^2-\MN^2}+{{1\over 2}\Gamma_4\over p^2-\MN^2}.
\end{eqnarray}
Noting that the $\pi NN$ coupling constant, $g$, is related to the invariant vertex functions as
\begin{eqnarray}\label{gGamma}
g=-\Gamma_1+\Gamma_2+2\Gamma_3
-\left(2-{m_\pi^2\over2\MN^2}\right)\Gamma_4\Bigg|_{p^2=\MN^2,(p-k)^2=\MN^2},
\end{eqnarray}
we clearly see that the coefficients of the double poles at $p^2=\MN^2$ and $(p-k)^2=\MN^2$ are proportional to $g$, i.e. $-{m_\pi^2\over 2}g$, $0$, $\MN^2g$ and $-\MN^2g$ for the structures $i\gamma_5$, $i\gamma_5\gsl p$, $i\gamma_5\gsl k$ and $\gamma_5\sigma_{\mu\nu}p^\mu k^\nu$, respectively,
but that the coefficients of the single poles at $p^2=\MN^2$ or $(p-k)^2=\MN^2$ are not determined by $g$.

In the chiral limit with $k=0$ the invariant correlation functions can be regarded as functions of $p^2$ and the discontinuities of the invariant correlation functions across the real $p^2$ axis are given by
\begin{eqnarray*}
{\rm Im}\Pi_1
&=&\pi\delta(p^2-\MN^2)\Gamma_1(\MN^2)-{{\rm P}\over p^2-\MN^2}{\rm Im}\Gamma_1(p^2),\cr
{\rm Im}\Pi_2
&=&\pi\delta(p^2-\MN^2)\Gamma_2(\MN^2)-{{\rm P}\over p^2-\MN^2}{\rm Im}\Gamma_2(p^2)\,\cr
{\rm Im}\Pi_3
&=&\pi\delta'(p^2-\MN^2)\MN^2g_0-\pi\delta(p^2-\MN^2)\MN^2{\rm Re}\Gamma_A'(\MN^2)+{{\rm Pf}\over (p^2-\MN^2)^2}\MN^2{\rm Im}\Gamma_A(p^2),\cr
{\rm Im}\Pi_4
&=&-\pi\delta'(p^2-\MN^2)\MN^2g_0+\pi\delta(p^2-\MN^2)\MN^2{\rm Re}\Gamma_B'(\MN^2)-{{\rm Pf}\over (p^2-\MN^2)^2}\MN^2{\rm Im}\Gamma_B(p^2),
\end{eqnarray*}
where
\begin{eqnarray*}
  &&\Gamma_A=-\Gamma_1+{p^2\over \MN^2}\Gamma_2+\left(1+{p^2\over \MN^2}\right)\Gamma_3-2{p^2\over \MN^2}\Gamma_4,\cr
  &&\Gamma_B=-\Gamma_1+\Gamma_2+2\Gamma_3-\left(1+{p^2\over \MN^2}\right)\Gamma_4,
\end{eqnarray*}
and $g_0$ denotes the $\pi NN$ coupling constant in the chiral limit.
For the structure $i\gamma_5$ the double pole vanishes,
 while for the structures $i\gamma_5\gsl k$ and $\gamma_5\sigma_{\mu\nu}p^\mu k^\nu$ the double poles survive but $i\gamma_5\gsl k$ and $\gamma_5\sigma_{\mu\nu}p^\mu k^\nu$ themselves vanish.
The coefficients of the single poles for the structures $i\gamma_5\gsl k$ and $\gamma_5\sigma_{\mu\nu}p^\mu k^\nu$ involve the derivatives of the invariant vertex functions with respect to $p^2$ and are therefore ill defined due to \Eq.{ReVF}.
The third term of ${\rm Im}\Pi_3$ as well as ${\rm Im}\Pi_4$ behaves as $\theta(p^2-\MN^2)(p^2-\MN^2)^{-1}$ 
since the absorptive part of the vertex function behaves as $p^2-\MN^2$ due to \Eq.{ImVF}, whose integral is also ill defined. 
We should note that the total dispersion integral is well defined, 
but when it is split into two parts as in \Eq.{TwoParts} each part is ill defined.

If one employs the effective lagrangian and calculates the vertex function at the tree level, one obtains $\Gamma_1=-g$ and $\Gamma_2=\Gamma_3=\Gamma_4=0$ for the pseudoscalar coupling scheme, while $\Gamma_3={g\over2}$ and $\Gamma_1=\Gamma_2=\Gamma_4=0$ for the pseudovector coupling scheme. 
If one substitutes these invariant vertex functions in \Eq.{ICFs}, one finds that two coupling schemes give the same result only for ${\rm Im}\Pi_4$.
This observation made Kim et al. in  Ref.~[\ref{KLO}] claim that only the structure $\gamma_5\sigma_{\mu\nu}p^\mu k^\nu$ is independent of the coupling schemes.
(They also claimed that the structure has less uncertainty in the OPE.)

The tree level calculation is meaningful for the double-pole term because the effect of the loops can be absorbed by the renormalization of the coupling constant.
But the tree-level calculation is not meaningful for the single-pole and continuum terms because the effect of the loops drastically change the structure.
In fact, at the one-loop level the second and third terms of ${\rm Im}\Pi_3$ and ${\rm Im}\Pi_4$ are generated which behave as expected by the general consideration above.

Another point worth while mentioning is the contribution of the excited baryon which couples to the nucleon-pion channel.
It is certainly true that the effective lagrangian which describes such a coupling generates a finite single-pole term.
But it is just a part of the contribution and it alone does not provide us with the structure expected from the general principle.
Thus, just adding the tree level contribution of the effective lagrangian for the excited baryon is insufficient.

Therefore, in our opinion a reasonable procedure to make an ansatz for the absorptive part of the invariant correlation function to be used in the sum rule would be first to transform it, for instance by multiplying by $p^2-\MN^2$, to a form which does not include ill-defined terms and then to parameterize it in accordance with the structure expected by the general principle, not by the tree level of the effective lagrangian.

Another way to construct a well-defined sum rule, a projrcted correlation function approach, is proposed in a separate paper~[\ref{preparation}].

In summary, we have reinvestigated the QCD sum rule for the $\pi NN$ coupling constant, $g$, starting from the vacuum-to-pion matrix element of the correlation function of the interpolating fields of two baryons.
We have studied in detail the physical content of the correlation function without referring to the effective lagrangian.
We have considered the invariant correlation functions by splitting the correlation function into different Dirac structures.
We have shown that the coefficients of the double poles are proportional to $g$
but that the coefficients of the single poles are not determined by $g$.
In the chiral limit the double-pole terms survive only for the Dirac structures $i\gamma_5\gsl k$ and $\gamma_5\sigma_{\mu\nu}p^\mu k^\nu$.
For these structures the single-pole terms as well as the continuum terms turn out to be ill defined in the dispersion integral.
Therefore, the use of naive QCD sum rules for these structures in the chiral limit is not justified.
A possible procedure to avoid this difficulty has been discussed.\\

\end{document}